\documentclass[12pt]{JHEP3}
\usepackage{amsmath,amssymb,bbm}
 \usepackage{epsf}
 \usepackage{epsfig}
 \usepackage{graphics}
\def\de{\partial}

\def\2{\frac12}
\def\4{\frac14}

\def\a{\alpha}
\def\b{\beta}
\def\g{\gamma}

\def\d{\delta}
\def\e{\epsilon}

\def\L{\Lambda}
\def\m{\mu}

\def\S{\Sigma}
\def\t{\tau}

\def\de{\partial}

\def\be{\begin{equation}}
\def\ee{\end{equation}}
\def\bea{\begin{eqnarray}}
\def\eea{\end{eqnarray}}

\author{Eric A. Bergshoeff, Mees de Roo, Sven F. Kerstan\\Centre for
Theoretical Physics, University of
Groningen, Nijenborgh 4, 9747 AG Groningen, The Netherlands\\
\email{E.A.Bergshoeff@rug.nl, M.de.Roo@rug.nl, s.f.kerstan.99@cantab.net}}
\author{Tom\'as Ort\'\i n\\Instituto de F\'\i sica Te\'orica
UAM/CSIC, Facultad de Ciencias C-XVI, C.U. Cantoblanco,
E-28049-Madrid, Spain\\ \email{Tomas.Ortin@cern.ch} }
\author{Fabio Riccioni
\\ Department of Mathematics, King's College, London WC2R 2LS, UK\\
\email{F.Riccioni@damtp.cam.ac.uk}}

\abstract{We give a universal
$SL(2,\mathbb{R})$--invariant expression for
all IIB $p$--brane actions with $p=-1,1,3,5,7,9$. The
Wess-Zumino terms in the brane actions are determined by
requiring (i)
target space gauge invariance and  (ii)
the presence of a single Born-Infeld vector. We find that
for $p=7$ $(p=9)$
brane actions with these properties only exist for orbits that
contain the standard D7--brane
(D9--brane). We comment about the actions for the other orbits.}

\title{$SL(2,\mathbb{R})$--invariant IIB Brane Actions }

\preprint{\hepth{0611036}\\
UG-06-08\\
IFT-UAM/CSIC-06-50\\
KCL-MTH-06-11}


\begin{document}

\section{Introduction\label{Intro}}

It is well--known that the duality group of the classical
IIB string theory is
$SL(2,\mathbb{R})$ and that this group of  duality transformations
gets broken to $SL(2,\mathbb{Z})$ at the
quantum level. At the level of the low energy limit of IIB
string theory the $SL(2,\mathbb{R})$  symmetry manifests itself as
a (non--linear) symmetry that acts on the fields of the IIB
supergravity multiplet \cite{SW,schwarz,HW}. In particular, the two scalars (
the dilaton and the axion) parametrize the coset
$SL(2,\mathbb{R})/SO(2) \equiv SU(1,1)/U(1)$.

When the Dp--branes of IIB supergravity were discovered
\cite{Polchinski:1995mt} a somewhat
unsatisfactory situation arose: the formulations of the worldvolume
actions for the Dp--branes broke the $SL(2,\mathbb{R})$ symmetry of the
theory. This applies for instance to the actions of
\cite{cederwall,superdbranes,schwarz2}. For special cases there have
been attempts to rectify this situation. For instance, an
$SL(2,\mathbb{R})$--invariant formulation of $(p,q)$--strings
\cite{pqstrings} has been given \cite{nokappa} even including
kappa--symmetry \cite{kappa}. This formulation made use of the fact
that in two spacetime dimensions the Born-Infeld vector is
equivalent to an integration constant describing the tension of a
string. Similarly, the case of 3--branes has been discussed
\cite{Cederwall}. In this case one makes use of the fact that in 4
spacetime dimensions the electric--magnetic dual of a Born--Infeld
vector is again a vector. Such special properties do not occur for
the branes with $p>3$ and indeed constructing an
$SL(2,\mathbb{R})$--invariant formulation of 5--branes turns out to
be problematic \cite{westerberg}.

In this paper we will fill this gap and provide an
$SL(2,\mathbb{R})$--invariant expression for all the branes of IIB
string theory. In doing this we make crucial use of the fact that only
recently  the supersymmetry and gauge transformations
for all $p$--form fields compatible with the IIB algebra have been  derived
\cite{IIBrevisited}\footnote{The same
has been done for the IIA case \cite{IIArevisited}.}.
These fields are a doublet of 2--forms, a singlet
4--form, a doublet of 6--forms, a triplet of 8--forms,
a quadruplet of 10--forms and a doublet of 10--forms:
\begin{equation}
A_{(2)}^\alpha\,,\hskip .3truecm A_{(4)}\,, \hskip .3truecm A_{(6)}^\alpha\,,
\hskip .3truecm A_{(8)}^{\alpha\beta}\,, \hskip .3truecm
A_{(10)}^{\alpha\beta\gamma}\,, \hskip .3truecm A_{(10)}^{\alpha}\,.
\end{equation}
Here we have used the $SU(1,1)$ notation with $\alpha=1,2$\footnote{We use
both (complex) $SU(1,1)$ and (real) $SL(2,\mathbb{R})$ notation.
The connection between
the two is explained in Appendix A which also contains our further
conventions.}.
In \cite{IIArevisited} the
gauge transformations and supersymmetries of these $p$--form fields
were given in a manifestly
$SU(1,1)$-invariant form.

These results  opened up the possibility of formulating, for all $p$,
$p$-brane actions in an
$SL(2,\mathbb{R})$-invariant way.
A first step
in this direction was taken in \cite{IIBbranes}, where
all possible branes for IIB were classified and
their tensions determined in an $SL(2,\mathbb{R})$-covariant way.
In particular, it was found that the D7--brane and D9--brane
belong to a {\sl nonlinear} doublet of $SL(2,\mathbb{R})$.
This is unlike the $(p,q)$--strings that form a linear doublet \cite{pqstrings}.

In this paper we continue the construction of the brane actions by
including the Born-Infeld worldvolume vector. This vector is
part of a doublet of vectors:
\begin{equation}\label{wvv}
V_{(1)}^\alpha\,,
\end{equation}
where the existence of the two different worldvolume vectors corresponds to
the fact that
either an F-string or a D-string (or, more generally, a $(p,q)$--string)
can end on the brane.
The challenge is now to construct a WZ term that at the same time
involves a single worldvolume vector and $p$--form fields that are
in non-trivial representations of $SU(1,1)$. In particular,
at first sight the triplet of 8--forms and the quadruplet of 10--forms
suggest that we introduce corresponding charges that transform as
a triplet $q_{\alpha\beta}$ and quadruplet $q_{\alpha\beta\gamma}$
of $SU(1,1)$, respectively.
Assuming that the worldvolume vector that occurs on the brane is given
by the combination $q_\alpha V_{(1)}^\alpha$ for certain constants $q_\alpha$
we will show in this
paper that, given certain requirements, a Wess--Zumino (WZ) term can only be
constructed for the restricted charges given by
\begin{equation}\label{restricted}
q_{\alpha\beta} = q_\alpha q_\beta\,,\hskip 3truecm q_{\alpha\beta\gamma} =
q_\alpha q_\beta q_\gamma\,.
\end{equation}
These charges include those of the standard D7--brane and D9--brane.
Note that, in the case of the D7--brane, the above restriction is
equivalent to the condition that
\begin{equation}
\det\, q_{\alpha\beta} = 0\,.
\end{equation}
For charges that belong to the other conjugacy classes of $SL(2,\mathbb{R})$,
with $\det\, (q_{\a\b}) \ne 0$, it is not possible to construct a
brane action of the required form.

In this paper we will derive a universal formula for the WZ term valid for all branes.
Furthermore, given the above restrictions on the charges we will show that
the brane tension that occurs in the kinetic terms of all brane actions
can also be given by an elegant universal formula. In this way
we obtain a unified expression of all $p$--brane actions. These brane  actions
are $SU(1,1)$--invariant provided we also rotate  the
constants $q_\alpha$ at the same time we rotate the worldvolume and
target space fields. In other words, we propose an
$SU(1,1)$-covariant family of actions.

This paper is organized as follows. In section \ref{actions} we derive the
general $SU(1,1)$--invariant
expression for all IIB brane actions. We will do this first for the
Wess-Zumino terms and, next, for the kinetic terms. Special cases
will be discussed in section \ref{specialcases} where we will
compare with other results in the literature.
We give our conclusions in
section \ref{conclusion}.
In appendix \ref{conventions} we give some of
 our conventions
and in appendix \ref{transformations} we list the
gauge transformations and invariant field strengths
of the different IIB $p$--form gauge potentials. 

\section{$SL(2,\mathbb{R}$)--invariant IIB Brane Actions} \label{actions}

The standard branes of IIB string theory, that is a doublet of
strings, a singlet of three-branes, a doublet of five-branes, a
nonlinear doublet of 7-branes and a non-linear doublet of 9-branes
\cite{IIBbranes}, all carry a world-volume vector-field. The need
for this can easily be seen by counting (bosonic and fermionic)
worldvolume degrees of freedom and requiring supersymmetry. The fact
that only one vector field is involved is related to the fact that
only one type of string can end on these branes. Since this string
belongs to a doublet of strings it is natural to introduce  an
$SU(1,1)$ doublet of worldvolume vectors $V_{(1)}^\alpha$, see
eq.~\eqref{wvv}, and next require that only a particular combination
occurs on the brane. Motivated by the case of D--branes we define a
gauge--invariant field strength  $\mathcal{F}^\a_{(2)}$ as
follows\footnote{In this paper we use the notation of
\cite{IIArevisited}, denoting $n$-forms $F_{\m_1 \dots \m_n}$ by
$F_{(n)}$. Antisymmetrization (with weight one) of the indices is
always understood.}:
\bea
\label{wvcurvature}
\mathcal{F}_{(2)}^\a =
F_{(2)}^\a + A_{(2)}^\a\,,\hskip 3truecm F_{(2)}^\a &=& 2 \partial
V_{(1)}^\a\,,
\eea
where $A_{(2)}^\a$ denotes the pull-back of the
target space 2--form field. Whenever that does not cause confusion
we will use the same symbol for the target space fields and their
pull-backs. In particular, we do not indicate the worldvolume scalars
that are involved in the pull-backs.
The worldvolume curvature \eqref{wvcurvature} is
invariant under the gauge transformations
\begin{equation}
\d_g V_{(1)}^\a = \partial \S^\a - \L_{(1)}^\a\,,\hskip 2truecm
\delta_g A_{(2)}^\a = 2\partial \Lambda_{(1)}^\a\,,
\end{equation}
where $\S^\a$ is the worldvolume gauge parameter and
$\L^\a_{(1)}$ is the (pull-back of the) gauge parameter
of the target space two form $A^\a_{(2)}$.

To characterize which type of string ends on the brane we require that
only the combination $q_\alpha V_{(1)}^\alpha$ occurs on the brane.
In the following two subsections we will derive expressions for
the WZ terms and for the kinetic terms.

\subsection{Wess--Zumino Terms}

The WZ-term of the brane actions are determined  by writing
down the most general Ansatz for a WZ-term and then demanding\footnote{
Note that the case $p=1$ is special since in 2 spacetime dimensions a vector
does not carry any worldvolume degree of freedom. Instead,
it  can be integrated out to yield an integration constant. Indeed,
the action of \cite{nokappa},
for instance, contains {\sl two} worldvolume vectors. The $p=1$ case will
be treated separately, see below.}
\begin{enumerate}
\label{criteria}
\item{Target Space Gauge invariance}
\item{A single worldvolume vector field ($p\ne 1$)}
\end{enumerate}
The second requirement is needed to have worldvolume supersymmetry.
As we already discussed the requirement of a single worldvolume
vector is satisfied by requiring that only the combination $q_\alpha
V_{(1)}^\alpha$ occurs on the brane. By convention we assume that
the case of D--branes, i.e., an F--string ending on the brane, is
covered by taking $q_{\a^\prime} =  (0,-1)$ 
where we work in the
$SL(2,\mathbb{R})$--basis\footnote{
To distinguish we use $\alpha=1,2$ in the $SU(1,1)$ basis and
$\alpha^\prime = 1,2$ in the  $SL(2,\mathbb{R})$ basis.}, 
see appendix A. All the other cases,
i.e., a general $(p,q)$--string ending on the brane, are then covered
by an $SL(2,\mathbb{R})$ transformation of the D--brane case and are
obtained by taking a general $q$--vector.

Our aim is to find a unified and $SU(1,1)$--invariant
WZ term for all IIB $p$-branes. In order to do this
it is useful to recall the universal formula for WZ terms in the
case of the (non $SU(1,1)$--invariant) D--branes:
\be\label{WZshort}
{\cal L}_{\rm WZ}(\text{D$p$-brane})  = C\, e^{\mathcal{F}_{(2)}}\,,
\ee
where $C$ is defined as the formal
 sum\footnote{Note that this sum also
contains a term involving $C_{(0)}\equiv \ell$ which is not required
by gauge invariance.} \bea C = \sum_n C_{(n)} = C_{(0)} + C_{(2)} +
C_{(4)} +C_{(6)} + C_{(8)} + C_{(10)}\,, \eea and $C_{(n)}$ are the
usual RR $n$--forms. It is understood here that after expanding the
exponential in eq.~\eqref{WZshort} in each term that particular
$C_{(n)}$ is chosen such that the product of forms adds up to a
$(p+1)$--form. Using this notation one can check that the WZ term is
invariant under the gauge transformations
\begin{equation}
\delta C = d \lambda + F_{(3)}\,\lambda\,,
\end{equation}
where $F_{(3)}$ is the curl of the NS-NS 2-form field $B$ and $\lambda$
is the formal sum
\bea
\lambda = \sum_n \lambda_{(n)} = \lambda_{(1)} + \lambda_{(3)}
+ \lambda_{(5)} + \lambda_{(7)}+ \lambda_{(9)}\,.
\eea
The $\lambda_{(n)}$ are the different RR gauge parameters.
To prove that the WZ term is gauge--invariant
requires a one line calculation where one uses that
$d\mathcal{F}_{(2)} = F_{(3)}$.
For this to work it is important that the gauge transformation of $C$
is  of the above form, i.e., it contains the usual
$d\Lambda$ term and terms that are all proportional to $F_{(3)}$.
This is not the case for the
$SU(1,1)$--covariant gauge potentials we have introduced in
\cite{IIBrevisited}. Their
gauge transformations are listed in Appendix B and it can be seen that
they contain $d\Lambda$ terms, terms proportional to $F_{(3)}^\a$ but also
additional terms proportional to $\Lambda_{(1)}^\a$.

Inspired
by the case of D--branes we make the following field redefinitions to
remove the extra terms from the gauge transformations:
\bea\label{redefinitions}
{\cal C}_{(2)}^\a &=& A_{(2)}^\a\,, \\
{\cal C}_{(4)}&=& A_{(4)} -\tfrac{3}{8} \tilde{q}_\a q_\b A_{(2)}^\a A_{(2)}^\b\,,\\
{\cal C}_{(6)}^\a &=& A_{(6)}^\a + 20 A_{(4)} A_{(2)}^\a - \tfrac{15}{2}
q_\b \tilde{q}_\g A_{(2)}^\a A_{(2)}^\b A_{(2)}^\g\,,\\
{\cal C}_{(8)}^{\a\b} &=& A_{(8)}^{\a\b} + \tfrac{7}{4} A_{(6)}^{(\a}
A_{(2)}^{\b )} + 35 A_{(4)}
  A_{(2)}^{\a} A_{(2)}^{\b} - \tfrac{105}{8}  q_\g \tilde{q}_\d
  A_{(2)}^\a A_{(2)}^\b A_{(2)}^\g A_{(2)}^\d\,, \\
{\cal C}_{(10)}^{\a\b\g} &=& A_{(10)}^{\a\b\g} -3 A_{(8)}^{(\a\b}
A_{(2)}^{\g)} -
  \tfrac{21}{4} A_{(6)}^{(\a} A_{(2)}^\b A_{(2)}^{\g)} -
  105 A_{(4)} A_{(2)}^\a A_{(2)}^\b A_{(2)}^\g \nonumber \\
  &+&
  \tfrac{315}{8} q_\d \tilde{q}_\e A_{(2)}^\a A_{(2)}^\b A_{(2)}^\g
  A_{(2)}^\d A_{(2)}^\e\,,
  \eea
where ${\tilde q}_\a$ is another doublet that satisfies
\be
\tilde{q}_{[\a} q_{\b]} = \tfrac{i}{2}\e_{\a\b}\,.
\label{normalisation}
\ee
We choose a basis such that for the case of D--branes
we have ${\tilde q}_{\a'}=(1,0)$ and $q_{\a'} = (0,-1)$.
Note that in this case $\tilde{q}$ is the S-dual of $q$,
see appendix \ref{conventions}.

We have not included the doublet of 10--forms since they
do not seem to fit in this family of potentials and require a
separate discussion, see below. After these redefinitions we end up
with the desired form of the gauge transformations:
  \bea
  \delta_g \mathcal{C} &=& d\Lambda +  F_{(3)}\
  \Lambda\,,\label{Cgaugetransf}
  \eea
where $F_{(3)} = 3 \de {\cal C}_{(2)}$ and $\mathcal{C}, \Lambda$
are defined by the formal sums\footnote{Note that in the expression
for ${\cal C}$ the first term in the sum, which will be discussed at
the end of this subsection, is not required by gauge--invariance of
the WZ term.}
\bea
\label{formalsum}
\mathcal{C} &=& \sum_{n,\a}
{\cal C}_{(n)}^{(\a)} = {\cal C}_{(0)} + {\cal C}_{(2)}^\a + {\cal
C}_{(4)} + {\cal C}_{(6)}^\a
+ {\cal C}_{(8)}^{\a\b}+{\cal C}_{(10)}^{\a\b\g}\,,\\
\Lambda &=& \sum_{n,\a}\Lambda^{(\a)}_{(n)} = \Lambda^\a_{(1)} +
\Lambda_{(3)} + \Lambda_{(5)}^\a + \Lambda_{(7)}^{\a\b} +
\Lambda_{(9)}^{\a\b\g}\,.
\eea
Eq. (\ref{Cgaugetransf}) applies to all the forms of rank higher
than 4, while for ${\cal C}_{(4)}$ it has to be replaced by
\be
  \d {\cal C}_{(4)} =4 \de {\L}_{(3)} + \tfrac{1}{2}
  q_\a F_{(3)}^\a \tilde{q}_\b \L_{(1)}^\b \quad .
\ee
The notation in (\ref{Cgaugetransf}) indicates that all $SU(1,1)$
indices in the second $\Lambda$ term are symmetrised with the $\a$
index of $F$, and all the terms have the same rank and the same
number of $SU(1,1)$ indices. Special cases are worked out in detail 
in Section \ref{specialcases}.

We find that in the new basis the WZ term can be cast into the
following universal form \be {\cal L}_{\rm WZ}(\text{$p$-brane}) = q
\cdot \mathcal{C}\, e^{q  \mathcal{F}_{(2)}}\,, \label{universalWZ}
\ee where $q \mathcal{F}_{(2)}$ stands for
$q_\a\mathcal{F}^\a_{(2)}$ and $q\cdot \mathcal{C}$ denotes
contraction of all $SU(1,1)$ indices of the
 ${\cal C}_{(n)}$ with as many $q$'s as are required, except for  ${\cal C}_{(2)}$ that must
be contracted with  $\tilde q$. Therefore, all terms in the WZ term
\eqref{universalWZ} are either independent of $\tilde q$ or at most
linear in $\tilde q$ where $\tilde q$ always occurs in the
combination ${\tilde q}_\a {\cal C}_{(2)}^\a$. Note that the formula
\eqref{universalWZ} implies, as anticipated in the introduction,
that in the leading term of the WZ terms the restrictions on the
charges given in \eqref{restricted} hold. We will show in subsection
(3.4) that without these restrictions on the charges it is not
possible to construct a WZ term that satisfies the criteria given in
\eqref{criteria}.

We close this subsection with two comments. First, we have not yet
specified the first term ${\cal C}_{(0)}$ in the formal sum
$\mathcal{C}$. This term leads to an expression in the WZ term that
is gauge--invariant by itself. For D$p$-branes with $p$ odd this
expression is given by $\ell\, \mathcal{F}_{(2)}^{(p+1)/2}$. To
reproduce this expression we must take
\be \label{C0}
{\cal C}_{(0)} =
-\frac{q_\a{\tilde q}_\b \mathcal{M}^{\a\b}}{q_\g q_\d
\mathcal{M}^{\g\d}}\,,
\ee
where the matrix ${\cal M}$ is given, in the $SL(2,\mathbb{R})$ basis,
in (\ref{scalarmatrix}).
Secondly, the doublet of 10--forms
$A_{(10)}^\a$  is not included by the universal formula
\eqref{universalWZ}. The reason is that the construction of a WZ
term for these 10--forms does not require the introduction of a
worldvolume vector field. This is due to the fact that their gauge
transformations only contain the leading term $\d A_{(10)}^\a =
d\Lambda_{(10)}^\a$ and therefore a WZ term of the form \be q_\a
A_{(10)}^\a \ee is already gauge--invariant by itself. However,
without a worldvolume vector it is not clear how to obtain an equal
number of bosonic and fermionic  worldvolume degrees of freedom and
establish worldvolume supersymmetry. We will not consider this case further
in this paper.

\subsection{Kinetic terms}

We next discuss the construction of the kinetic terms. It is convenient to
work in Einstein frame, since the metric $g_E$ is $SU(1,1)$--invariant.
The action for a $p$-brane in Einstein frame can be written as
\be
{\cal L}_{\text{kinetic}}(\,p\text{--brane}) =  \t_{p,E}\, \sqrt{\det\left(g_E + s\, q \mathcal{F} \right)}\,,
\ee
where $\t_{p,E}$ is the brane tension for the $p$-brane in Einstein frame and $s$ is a
scalar function of the scalar fields. Using the expressions for the tensions,
already obtained in \cite{IIBbranes} from supersymmetry, together
with our universal formula for the WZ term \eqref{universalWZ}
we find the following general formula for the $p$--brane tensions\footnote{
For the $p=1$ case, see subsection (3.2).}:
\bea\label{tension1}
\t_{p,E} &=& \bigl (q q \mathcal{M}\bigr)^{\tfrac{p-3}{4}}
\hskip 2.75truecm p\ne 1\,,\\
\t_{1,E} &=& \bigl(\tilde q\tilde q \mathcal{M}\bigr)^{\tfrac{1}{2}}
\hskip 3.15truecm p=1\,,
\eea
where $q q \mathcal{M}$ stands for $q_\a q_\b \mathcal{M}^{\a\b}$.

Further, we find we can
write the kinetic terms as
\be
{\cal L}_{\text{kinetic}}(\,p\text{--brane})=
\t_{p,E}\, \sqrt{\det{\left(g_E + \frac{q \mathcal{F}}{(q q \mathcal{M})^{1/2}}\right)}}\,.
\ee
Summarizing, we find that the SU(1,1)--invariant Lagrangian for general
IIB $p$--branes is given by ($p \ne1$)
\be\label{finalpbraneaction}
\boxed{
{\cal L}(\,p\text{--brane})\ = \
\t_{p,E}\, \sqrt{\det{\left(g_E + \frac{q \mathcal{F}}{(q q\mathcal{M})^{1/2}}\right)}}
\ +\
q \cdot \mathcal{C}\, e^{q \mathcal{F}_{(2)}}
}\,,
\ee
with the tension given by eq.~\eqref{tension1}, the
scalar matrix $\cal M$ given by
\eqref{scalarmatrix} and the formal sum $\mathcal{C}$ defined in \eqref{formalsum}.
The case $p=1$ is special in the sense that in that case the
worldvolume vectors can be integrated away, see subsection (3.2).

Note that all factors of $q_\a,{\tilde q}_\a$ in \eqref{finalpbraneaction}
are such that, if we assign a dimension $\Delta = 1, \Delta=-1$ to
$q_\a$ and ${\tilde q}_\a$, respectively, all terms in a given $p$--brane
action have dimension $\Delta = \tfrac{1}{2}(p-3)$.

\section{Special cases} \label{specialcases}

In this section we give explicit details for special cases and compare with the
literature.

\subsection{(--1)--branes}

This case corresponds to the orbit of D-instantons and is special in the
sense that we now work with Euclidean IIB supergravity. The $SU(1,1)$--invariant
instanton action is given by
\be
{\cal L}_{\text{(-1)--brane}} = (q  q \mathcal{M})^{-1}\ - \
\frac{q{\tilde q}\mathcal{M}}{q q \mathcal{M}}\,.
\ee
For $q_{\a^\prime}=  (0,-1)$ and ${\tilde q}_{\a^\prime} = (1,0)$
 we recover the standard D-instanton Lagrangian
${\cal L}_{\text{ D--instanton}} \sim e^{-\phi}+\ell$.

\subsection{1--branes}
The brane action for strings is \cite{pqstrings}
\be
{\cal L}_{\text{1--brane}} =
(\tilde q\tilde q\mathcal{M})^{1/2}\, \sqrt{\det{g_E }}\ +\
\tilde{q}_\a {\cal C}_{(2)}^\a.
\ee
We use here a form of the Lagrangian where there are no
Born-Infeld vectors.
Unlike all other
branes, we use ${\tilde q}_\a$ instead of $q_\a$ for the leading
term in the WZ terms. This fits with the fact that the dimension
of the Lagrangian for strings is given by $\Delta=-1$. The constants
$q_\a$ are absent.
Note that the construction of a $p=1$
gauge--invariant WZ term does not require the introduction of a
worldvolume vector, unlike the $p>1$ branes. The $p=1$ Lagrangian is
equivalent to the one given in \cite{nokappa} if we identify
${\tilde q}_{\a^\prime} = (p,q)$ with the two integration constants that
follow from integrating out the two worldvolume vectors that occur
in the formulation of \cite{nokappa}.

\subsection{3-branes}
The brane action for the $p=3$ case is given by
\be
{\cal L}_{\text{3--brane}} = \sqrt{\det{\left(g_E + \frac{q \mathcal{F}}{(
qq\mathcal{M})^{1/2}}\right)}}\ +\ {\cal C}_{(4)}  + \tfrac{3}{4}
\tilde{q}_\a q_\b {\cal C}_{(2)}^\a {\cal F}_{(2)}^\b
+ {\cal C}_{(0)} (q{\cal F}_{(2)})^2 \,,\label{WZ4}
\ee
with ${\cal C}_{(0)}$ defined in \eqref{C0}.
This is precisely the action that one obtains by dimensional
reduction of the PST action \cite{Pasti:1995tn,Pasti:1996vs} 
for a self-dual tensor in six dimensions
\cite{Berman:1997iz,Nurmagambetov:1998gp,Berman:1998sz}.
It is interesting to apply an electric--magnetic duality
transformation to the worldvolume vector $q_\a V_{(1)}^\a$ and to
compare with \cite{Cederwall}. We find that after an
electric--magnetic duality transformation we end up with the same
action but with the electric potential $q_\a V_{(1)}^\a$ replaced by
a magnetic one, say $M_{(1)}$, and with everywhere else $q_\a$
replaced by ${\tilde q}_\a$. On the other hand, in our 
basis (\ref{choices})
the effect of an S--duality transformation is to replace $q_\a$ by
${\tilde q}_\a$ everywhere, including the term $q_\a V_{(1)}^\a$.
Identifying
\be 
M_{(1)} = {\tilde q}_\a V_{(1)}^\a 
\ee 
we see that the
two operations coincide, i.e., an S--duality acts on the worldvolume
vector like an electric--magnetic (Hodge) duality transformation. 
This agrees with \cite{Cederwall}.

\subsection{5-branes}
The action for 5--branes is given by
\bea \label{fivebrane}
{\cal L}_{\text{5--brane}} &=&
(q q\mathcal{M})^{1/2}
\sqrt{\det{\left(g_E + \frac{q \mathcal{F}}{(q q\mathcal{M})^{1/2}}\right)}}  \label{WZ6}\\
&+&  q_\a \left( {\cal C}^\a_{(6)} - 60 {\cal C}_{(4)}{\cal
F}^\a_{(2)} - \tfrac{45}{2} \tilde{q}_\b q_\g  {\cal
C}^\b_{(2)}{\cal F}^\g_{(2)} {\cal F}^\a_{(2)}\right)
+ {\cal C}_{(0)} (q{\cal F}_{(2)})^3\nonumber
 \,.
\eea
 Gauge invariance of the WZ term implies that the 6-form
transforms as
  \be
  \delta_g {\cal C}_{(6)}^\a = 6 \de {\L}_{(5)}^\a - 80 F_{(3)}^\a {\L}_{(3)}
  \quad ,
  \ee
which is indeed like in eq. (\ref{Cgaugetransf}).

A different attempt to construct an $SU(1,1)$--invariant 5--brane
action was undertaken in \cite{westerberg}. Although the formula of
\cite{westerberg} is not complete, it would be interesting to see
whether there is any relation between the result of \cite{westerberg}
and \eqref{fivebrane}.

\subsection{7-branes}
The case of 7-branes is more subtle due to two reasons. First of all there
are different conjugacy classes of 7-branes solutions which are
distinguished by the value of $\det\, (q_{\a\b})$. Secondly,
to define 7--branes globally, one needs to consider other
7--branes at different positions in space\footnote{For a careful discussion of the global properties
and the role of the three different conjugacy classes, see
\cite{inpreparation}.}. The
$\det\, (q_{\a\b})=0$
conjugacy class contains the D7--brane.
Ignoring global properties, i.e., restricting to the
dynamics of small fluctuations, this class has the following brane action:
\bea
{\cal L}_{\text{7--brane}} &=&
q q\mathcal{M} \,
\sqrt{\det{\left(g_E + \frac{q \mathcal{F}}{(q q\mathcal{M})^{1/2}}\right)}}\ \label{WZ8}\\
&+& \ q_\a q_\b \left[ {\cal C}^{\a\b}_{(8)} -  7 {\cal C}^\a_{(6)}
{\cal F}^\b_{(2)} + 210 {\cal C}_{(4)}   {\cal F}^\a_{(2)} {\cal
F}^\b_{(2)}
 + \tfrac{105}{2}  \tilde{q}_\g q_\d  {\cal C}^\g_{(2)} {\cal F}^\d_{(2)}
 {\cal F}^\a_{(2)} {\cal F}^\b_{(2)}\right]\nonumber\\
&+&\ {\cal C}_{(0)} (q{\cal F}_{(2)})^4
\,.\nonumber
\eea
The gauge transformation of the 8-form is
  \be
  \d {\cal C}_{(8)}^{\a\b} = 8  \de {\L}_{(7)}^{\a\b} -14 F_{(3)}^{(\a}
  {\L}_{(5)}^{\b)} \quad ,
  \ee
which again is of the form (\ref{Cgaugetransf}).

The above WZ-term has a 7-brane ``charge'' matrix $q_{\a\b} = q_\a
q_\b$. The determinant of this matrix is, by construction, zero (the
matrix has two linearly dependent columns). It is natural to ask
whether brane actions for the other conjugacy classes, i.e., with
$\det\, (q_{\a\b}) \ne 0$ can also be constructed. Assuming such a
charge matrix we write down the first few terms for the most general
ansatz for  a WZ-term:
\be WZ_{(8)} = q_{\a\b} \left[ A^{\a\b}_{(8)}
+ a_1 A^\a_{(6)} A^\b_{(2)} + a_2 A^\a_{(6)} F^\b_{(2)} +
\ldots\right] \label{WZ7alt}
\ee
where $a_1,a_2$ are to be
determined. Demanding that there be only one vector field on the
brane requires $a_2 = 0$. This can be seen by assuming $a_2\ne0$.
Then the second column of the matrix $q_{\a\b}$ must be zero,
because otherwise we would introduce two gauge fields in the
$a_2$-term. But the second column being trivial implies $\det\
(q_{\a\b}) = 0$, in contradiction to our assumption, and so we must
have $a_2=0$. We now take a look at the gauge transformation of
(\ref{WZ7alt}). The terms of the type $\partial A_{(6)}^\a
\L^\b_{(1)}$ and of the type $\partial A_{(2)}^\a \L^\b_{(5)}$, both
of which are produced by the $A^{\a\b}_{(8)}$ and the $A^\a_{(6)}
A^\b_{(2)}$--terms in our ansatz (remember that we already have
eliminated the last term in (\ref{WZ7alt}), which would also have
been a source of such terms), cannot be canceled at the same time
for any choice of $a_1$. This shows that it is not possible to
construct a brane action containg a single Born-Infeld vector,
for any 7-brane with $\det\, (q_{\a\b})\ne 0$.

Of course, the above analysis does not exclude non-standard brane actions.
For instance, recalling that the monodromy of a $\det\, (q_{\a\b})>0$ brane
can be obtained as the product of monodromies corresponding to
two  $\det\, (q_{\a\b})=0$ branes one could view a  $\det\, (q_{\a\b})>0$ brane
as a bound state of two  $\det\, (q_{\a\b})=0$ branes. This suggests that we
might consider a (non-Abelian) brane action containing two vector
fields\footnote{We thank Jelle Hartong
for a discussion on this possibility.}.

\subsection{9-branes}

Finally, we consider the case of 9-branes. The 9-branes related to
the D9-brane (the nonlinear doublet of 9-branes) have the following
brane action
\bea
{\cal L}_{\text{9--brane}} &=& (q
q\mathcal{M})^{3/2}\, \sqrt{\det{\left(g_E + \frac{q \mathcal{F}}{(
q  q \mathcal{M})^{1/2}}\right)}}\ \nonumber
\\
&+& \ q_\a q_\b q_\g \left[ {\cal C}^{\a\b\g}_{(10)} +  15 {\cal
C}^{\a\b}_{(8)} {\cal F}^\g_{(2)} -\tfrac{105}{2} {\cal C}^\a_{(6)}
{\cal F}^\b_{(2)} {\cal F}^\g_{(2)} \right.
\nonumber\\
&+ &\left. 1050 {\cal C}_{(4)} {\cal F}^\a_{(2)} {\cal F}^\b_{(2)}
{\cal F}^\g_{(2)} + \tfrac{1575}{8}  \tilde{q}_\d q_\e {\cal
C}^\d_{(2)} {\cal F}^\e_{(2)} {\cal F}^\a_{(2)} {\cal F}^\b_{(2)}
{\cal F}^\g_{(2)} \right]\nonumber\\
&+&\  {\cal C}_{(0)} (q{\cal F}_{(2)})^5
\,. \label{WZ10} \eea The WZ term is gauge
invariant provided that
  \be
  \d {\cal C}_{(10)}^{\a\b\g} = 10 \de {\L}_{(9)}^{\a\b\g} + 40
  F_{(3)}^{(\a}
  {\L}_{(7)}^{\b\g)} \quad ,
  \ee
which is of the form (\ref{Cgaugetransf}). Note that, unlike the case of
7--branes, the other conjugacy classes, not containing the D9--brane,
are not supersymmetric \cite{IIBbranes}.

\section{Conclusion} \label{conclusion}

In this paper we have presented an elegant $SU(1,1)$--invariant
expression for all $p$--brane
actions of the IIB theory, see eq.~\eqref{finalpbraneaction}.
We only considered the bosonic terms in the action. It is natural to
also consider the fermionic terms and require kappa--symmetry.
This requires a $SU(1,1)$--covariant superspace formulation
of IIB supergravity.

Concerning the 7--branes,
it would be
interesting to perform a zero mode analysis on the 7--brane solutions
for all conjugacy classes and
from that point understand why only the 7--brane solution belonging
to the $\det\, (q_{\a\b})=0$  conjugacy class has a single
vector field zero mode. Furthermore, one could then determine
what the zero modes are, if any, for the other conjugacy classes.

As far as the 9--branes are concerned, it remains unclear what the
interpretation is of the doublet of 10--form potentials. We have seen
that a gauge--invariant WZ term does not contain a worldvolume vector
field.
One would therefore expect that also
the kinetic term does not contain such a vector field.
Nevertheless, if
kappa--symmetry is going to work we expect to have 8 fermionic worldvolume
degrees of freedom like all the other branes and they need to be matched
by 8 bosonic degrees of freedom. In ten dimensions such bosonic
degrees of freedom can only be described by a vector belonging to a
vector multiplet.

Finally, to construct our central formula \eqref{finalpbraneaction}
it was crucial
to perform the field redefinitions \eqref{redefinitions}. These
field redefinitions involve the vectors $q_\a$ and ${\tilde q}_\a$.
It would be interesting to obtain a better understanding of these
redefinitions and of the role of the $SU(1,1)$--covariant
${\cal C}$--potentials in IIB string theory.

\section*{Acknowledgements}

We thank Jelle Hartong for useful discussions.
T.O.~and F.R.~would like to thank the University of Groningen for
hospitality. E.B., S.K., T.O.~and M. de R.~are supported by the
European Commission FP6 program MRTN-CT-2004-005104 in which E.B.,
S.K. and M. de R. are associated to Utrecht university and T.O.~is
associated to the IFT-UAM/CSIC in Madrid. The work of E.B.~and
T.O.~is partially supported by the Spanish grant BFM2003-01090. The
work of T.O.~has been partially supported by the Comunidad de Madrid
grant HEPHACOS P-ESP-00346. The work of F.R. is supported by the EU
contract MRTN-CT-2004-512194 and by the PPARC grant PP/C507145/1.

\appendix
\section{Conventions} \label{conventions}

We raise and lower $SU(1,1)$ indices with the two--dimensional Levi--Civita
tensor $\e$:
\begin{eqnarray}
q^\a &=& \e^{\a\b}q_\b\,,\hskip 4truecm
q_\b = q^\a \e_{\a\b}\,.
\end{eqnarray}
A $SU(1,1)$--doublet $q_\a$ satisfies the following reality condition:
\be
(q_1)^\star = q_2\,.
\ee
Instead of using the $SU(1,1)$--basis, with complex components $q_\a$, it is
sometimes convenient to use the $SL(2,\mathbb{R})$--notation with
real components $q_{\a^\prime}$. The two bases are related via the
following transformation:
\bea
q_{1^\prime} &=& \tfrac{1}{\sqrt 2}(q_1+q_2)\,, \hskip 3truecm
q_{2^\prime} = \tfrac{i}{\sqrt 2}(q_1 - q_2)\,.
\eea
With these conventions we have that $\e^{\a\b}q_\a r_\b =
\e^{\a^\prime\b^\prime}q_{\a^\prime} r_{\b^\prime}$ with $\epsilon^{12}=1$
and $\e^{1^\prime 2^\prime} = i$.
Note that under S--duality we have
\bea
q_1 &\stackrel{S}{\rightarrow} -iq_1\,,\hskip 4truecm
q_{1^\prime} \stackrel{S}{\rightarrow} -q_{2^\prime}\,,\\
q_2 &\stackrel{S}{\rightarrow} +iq_2\,,\hskip 4truecm
q_{2^\prime} \stackrel{S}{\rightarrow} +q_{1^\prime}\,.
\eea
In the text we have also defined a doublet ${\tilde q}_\a$ that
satisfies the relation \bea \tilde{q}_{[\a} q_{\b]} = \tfrac{i}{2}
\e_{\a\b}\,. \eea

We use an $SL(2,\mathbb{R})$--basis  where the case of D--branes is
recovered by making the choices:
\bea \label{choices}
{\tilde q}_{\a^\prime} = \begin{pmatrix}1\cr 0\end{pmatrix}\,,\hskip 3truecm
q_{\a^\prime} = \begin{pmatrix}0\cr -1\end{pmatrix}\,.
\eea
Then we have under S-duality
\bea
q_{\a^\prime} \stackrel{S}{\rightarrow} {\tilde q}_{\a^\prime}\,.
\eea
In our basis
the $2\times 2$ scalar matrix $\mathcal{M}$ is given
by
\be
\label{scalarmatrix}
{\cal M}^{\a^\prime\b^\prime}\ = \ e^\phi
\begin{pmatrix}
\ell^2 + e^{-2\phi} &\ell\cr
\ell & 1
\end{pmatrix}\,.
\ee

For the convenience of the reader we give
the value of some general $SU(1,1)$--invariant
expressions for the choices \eqref{choices}:
\bea
{\tilde q}_\a A_{(2)}^\a &\rightarrow& C_{(2)}\,,
\hskip 3truecm q_\a A_{(2)}^\a \rightarrow B_{(2)}\,,\\
{\tilde q}_\a A_{(n)}^\a &\rightarrow& B_{(n)}\,, \hskip 3truecm
q_\a A_{(n)}^\a \rightarrow C_{(n)}\,,\hskip 1truecm n\ne 2\,. \eea

\section{The $p$--Form Gauge Fields of IIB Supergravity} \label{transformations}
For convenience, we provide the gauge transformations and
field strengths for all IIB $p$--form gauge fields  as they were
determined in \cite{IIBrevisited}.

The $p$--form  gauge fields
of IIB supergravity are a singlet
4--form, a doublet of 2--forms, 6--forms and 10--forms, a triplet of 8--forms
and a quadruplet of 10--forms.
The gauge transformations of these gauge fields are:
\begin{eqnarray}
\d A^\a_{\m_1\m_2} &=& 2 \de_{[\m_1} \L^\a_{\m_2]}\quad ,
\\
\d A_{\m_1\ldots\m_4} &=& 4 \de_{[\m_1} \L_{\m_2\m_3\m_4]}
  - \tfrac{i}{4} \e_{\g\d} \L^\g_{[\m_1}F^\d_{\m_2\m_3\m_4]}\quad ,
\\
\d A^\a_{\m_1\ldots\m_6} &=& 6 \de_{[\m_1} \L^\a_{\m_2\ldots\m_6]}
     - 8 \L^\a_{[\m_1} F_{\m_2\ldots\m_6]}
  - \tfrac{160}{3} F^\a_{[\m_1\m_2\m_3}\L_{\m_4\m_5\m_6]}\quad ,
\\
\d A^{\a\b}_{\m_1\ldots\m_8} &=& 8 \de_{[\m_1}
\L^{\a\b}_{\m_2\ldots\m_8]}
     +\tfrac{1}{2} F^{(\a}_{[\m_1\ldots\m_7} \L^{\b)}_{\m_8]}
  - \tfrac{21}{2} F^{(\a}_{[\m_1\m_2\m_3} \L^{\b)}_{\m_4\ldots\m_8]}\quad ,
\\
\d A^{\a}_{\m_1 \ldots \m_{10}} &=& 10 \de_{[\m_1} \L^{\a}_{\m_2\ldots\m_{10}]}\quad ,
\\
\d A^{\a\b\g}_{\m_1 \ldots \m_{10}} &=&
   10 \de_{[\m_1} \L^{\a\b\g}_{\m_2\ldots\m_{10}]}
   -\tfrac{2}{3} F^{(\a\b}_{[\m_1 \ldots \m_9} \L^{\g)}_{\m_{10}]}
  + 32 F^{(\a}_{[\m_1 \m_2 \m_3} \L_{\m_4 \ldots \m_{10}]}^{\b\g)} \quad .
\end{eqnarray}
The expressions for the corresponding field strengths are given by:
\begin{eqnarray}
F^\a_{\mu_1\mu_2\mu_3} &=& 3 \de_{[\mu_1} A^\a_{\mu_2 \mu_3]} \quad ,
\\
F_{\mu_1 \ldots \mu_5} &=& 5 \de_{[\m_1} A_{\m_2\ldots\m_5 ]}
   + \tfrac{5i}{8} \e_{\a\b} A^\a_{[\m_1\m_2} F^{\b}_{\m_3\m_4\m_5 ]} \quad ,
\\
F^{\a}_{\mu_1 \ldots \mu_7} &=& 7 \de_{[\m_1} A^\a_{\m_2 \ldots \m_7]}
     + 28 A^\a_{[\m_1\m_2} F_{\m_3\ldots\m_7]}
  - \tfrac{280}{3} F^\a_{[\m_1\m_2\m_3}A_{\m_4\ldots\m_7]} \quad ,
\\
F^{\a\b}_{\mu_1 \ldots \mu_9} &=& 9 \de_{[\m_1} A^{\a\b}_{\m_2\ldots\m_9]} +
  \tfrac{9}{4} F^{(\a}_{[\m_1\ldots\m_7} A^{\b)}_{\m_8\m_9]}
 - \tfrac{63}{4} F^{(\a}_{[\m_1\m_2\m_3}A^{\b)}_{\m_4\ldots\m_9]} \quad ,
\\
F^{\a}_{\mu_1 \ldots \mu_{11}} &=&
 11 \partial_{[\mu_1} A^{\a}_{\mu_2 \ldots \mu_{11}]} = 0\quad ,
\\
F^{\a\b\g}_{\mu_1 \ldots \mu_{11}} &=&
  11 ( \partial_{[\mu_1} A^{\a\b\g}_{\mu_2 \ldots \mu_{11}]}
 -\tfrac{1}{3} F^{(\a \b}_{[\mu_1 \ldots \mu_9} A^{\g)}_{\mu_{10} \mu_{11}]}
+ 4  F^{(\a}_{[\mu_1 \mu_2 \mu_3} A^{\b \g)}_{\mu_4 \ldots \mu_{11}]}) = 0 \quad .
\end{eqnarray}
Note that all curvature terms that occur at the right-hand-side of the above equations
(both the gauge transformations and the expressions for the curvatures) are
related to the doublet of 2--form gauge fields, i.e. they are proportional to
$\Lambda_\mu^\alpha, A_{\mu\nu}^\alpha$ or $F_{\mu\nu\rho}^\alpha$.

\end{document}